\documentclass[a4paper]{article}
\usepackage{amssymb,graphicx,amsmath,amsfonts,wasysym,cite}
\usepackage[utf8]{inputenc}
\usepackage[margin=2.5cm]{geometry}

\title{Multi-particle quantum-statistical correlation functions in a Hubble-expanding hadron gas}

\author{M\'at\'e Csan\'ad$^1$, Antal Jakov\'ac$^1$, S\'andor L\"ok\"os$^{1,2}$,\\ Ayon Mukherjee$^1$, Srikanta Kumar Tripathy$^1$\\
$^1$: E{\"o}tv{\"o}s Lor{\'a}nd University, H-1117 Budapest, P{\'a}zm{\'a}ny P. s. 1/A, Hungary\\
$^2$: Eszterházy Károly University, Mátrai út 36, H-3200, Gyöngyös, Hungary}

\begin{document}

\maketitle

\begin{abstract}
Quantum-statistical correlation measurements in high-energy physics represent an important tool to obtain information about the space-time structure of the particle-emitting source. There are several final state effects which may modify the measured femtoscopic correlation functions. One of these may be the interaction of the investigated particles with the expanding hadron gas, consisting of the other final state particles. This may cause the trajectories -- and hence the phases -- of the quantum-correlated pairs to be modified compared to free streaming. The resulting effect and could be interpreted as an Aharonov--Bohm-like phenomenon, in the sense that the possible paths of a quantum-correlated pair represent a closed loop, with an internally present field caused by the hadron gas. In this paper, the possible role of the effect in heavy-ion experiments is presented with analytical calculations and a simple numerical model. The modification of the strength of multi-particle Bose-Einstein correlation functions is investigated, and the is found that in case of sufficiently large source density, this effect may play a non-negligible role.
\end{abstract}

\section{Introduction}
\label{intro}

Investigating particle correlations is a versatile tool often utilised in experimental particle and nuclear physics. In general, many different physical processes lead to correlated particle production: collective flow, jets, resonance decays, conservation laws. In high energy particle collisions, for identical bosons, the main source of correlations at low relative momenta is the Bose--Einstein (BE) quantum statistics~\cite{Goldhaber:1960sf,Baym:1997ce}, or in other words, the Hanbury Brown and Twiss (HBT) effect~\cite{HanburyBrown:1957rsa}. This is based on the indistinguishability of the two identical bosons and their symmetrical pair wave-function.~\cite{Glauber:1962tt} These discoveries led to the birth of femtoscopy~\cite{Lednicky:2001qv}, the goal of which field is to explore the femtometer-scale space-time geometry of the particle emitting source by measuring Bose-Einstein correlation functions of identical bosons (or Fermi-Dirac correlations of fermions). Besides quantum statistics, multiple phenomena affect the measured momentum correlations, among which the most important are final state interactions~\cite{Kincses:2019rug}.

\begin{figure}
\centering
\includegraphics[width=0.5\textwidth]{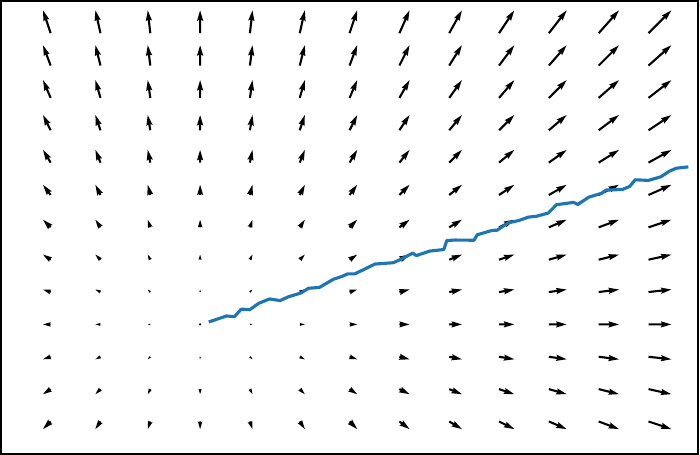}
\caption{\label{fig:ahbhubble} Illustration of the effect discussed in this manuscript, where the Hubble-flowing hadron gas is depicted with black arrows, along with a hypothetical particle path (not based on a calculation, just drawn as an illustration). The test particle moves in a Hubble-expanding hadron-gas and interactions with the other particles cause the movement of the particle to be modified. In reality, it is rather the velocity of the particle that is modified, not the path. That nevertheless causes a phase-shift, modifying quantum-statistical correlations.}
\end{figure}

Another effect that may be considered as modifying the momentum correlation functions is the interaction with the surrounding hadron gas. One may view this as an Aharonov--Bohm-like effect \cite{Aharonov:1959fk}, in the sense that the path of any given pair is a closed loop, as illustrated in Fig.~\ref{fig:ahbhubble}. The (electromagnetic, strong, etc.) fields inside of this closed path cause a phase shift in the pair wave function, proportional to the flux enclosed by the path. The phase shift, as discussed subsequently, modifies the effect of the quantum-statistical correlations. In this paper, the possible role of this effect in heavy-ion collision experiments is demonstrated. 

The paper is structured as follows: in Section \ref{s:becintro}, an introduction to Bose-Einstein correlations in high-energy physics is given and in Section \ref{s:lambda23}, the role of randomly fluctuating fields in two- and three-particle correlations is discussed. Then, in Section \ref{s:toymodel}, a simple model is set up to give quantitative details on how such a random field affects the Bose-Einstein correlation functions. Finally, Section \ref{s:results} concludes the discussion with the presentation of an observable that is sensitive to the aforementioned effect.

\section{Bose-Einstein correlations in high-energy physics}\label{s:becintro}

The two-particle momentum correlation function is defined, in general, as
\begin{align}
C_2(p_1,p_2)\equiv\frac{N_2(p_1,p_2)}{N_1(p_1)N_1(p_2)},\label{e:c2def}
\end{align}
where $N_1(p)$ and $N_2(p_1,p_2)$ are the one- and two-particle invariant momentum distributions, with $p_1$ and $p_2$ being the (four-)momenta. Neglecting final state interactions, the pair wave-function of  bosons is a symmetrised plane wave. This in turn leads~\cite{Yano:1978gk} to the expression of the correlation function by means of the phase-space density of the particle-emitting source $S(x,p)$ as
\begin{equation}
C_2(p_1,p_2)= 1+ {\rm Re}\frac{\widetilde S(q,p_1)\widetilde S^*(q,p_2)}{\widetilde S(0,p_1)\widetilde S^*(0,p_2)},\label{e:Cp1p2:general}
\end{equation}
where $q\equiv p_1-p_2$ is the relative momentum, complex conjugation is denoted by $^*$, and $\widetilde S(q,p)$ denotes the Fourier transform of the source:
\begin{equation}
\widetilde S(q,p)\equiv\int S(x,p) e^{iqx} d^4x.\label{e:tildeSdef}
\end{equation}

It is customary to introduce the average momentum $K\equiv 2(p_1+p_2)$, and then, based on the smoothness approximation~\cite{Lisa:2005dd}, in the kinematic domain of $p_1\approx p_2\approx K$ one obtains:
\begin{equation}
C_2(q,K)=1+\frac{|\widetilde S(q,K)|^2}{|\widetilde S(0,K)|^2}.
\label{e:C2expr}
\end{equation}
This equation essentially means, that the momentum correlations are connected to the source density $S(x,p)$, and hence by measuring these correlations, the femtometer-scale structure of the source can be investigated.

From Eq.~\eqref{e:C2expr} it is clear that the correlation function takes the value $2$ at zero relative momentum. However, experimentally $C_2(0) = 1+\lambda_2$, where $\lambda_2$ is the so-called intercept parameter (or the strength of the correlation function), and usually $\lambda_2 \le 1$ holds. The formula for the correlation function then may be empirically modified as
\begin{equation}
C_2(q,K)=1+\lambda_2\frac{|\widetilde S(q,K)|^2}{|\widetilde S(0,K)|^2}. \label{e:C2:corehalo}
\end{equation}
This empirical fact can be understood in terms of the core-halo model~\cite{Csorgo:1994in}. This treats the source as a sum of two components. One is the \emph{core}, which consists of primordial hadrons mainly, and the Fourier transform of this resolvable in momentum ($q$) space by the correlation measurement. The other component is a much wider\emph{halo}, consisting of decay products of long-lived
resonances (that travel much farther than $\simeq 10$ fm: $\eta$, $\eta'$, $\omega$, $K^0_S$, etc). The Fourier transform of the broad halo would be a very sharp peak at $q=0$, and this is experimentally essentially unresolvable. The intercept $\lambda_2$ (i.e. the extrapolation of the measured \emph{visible} correlation function to zero relative momentum) is in this picture obtained as the square of the fraction of pions coming from the core~\cite{Csorgo:1994in}:
\begin{equation}
\lambda_2 = f_c^2, \quad \textnormal{where} \quad f_c = \frac{N_{\rm core}}{N_{\rm core}+N_{\rm halo}} ~.
\label{e:lambda:corehalo}
\end{equation}
Hence $\lambda_2$, the strength of the two-particle correlation measures the fraction of primordial pions. This leads to an interesting application: the pair momentum ($K$) dependence of $\lambda_2$ may reveal a mass decrease of the ``prodigal'' $\eta'$ meson due to chiral U$_A(1)$ symmetry restoration~\cite{Kapusta:1995ww,Vance:1998wd,Csorgo:2009pa,Adare:2017vig}. The possible presence of partially coherent pion production distorts the above picture~\cite{Csorgo:1999sj,Sinyukov:1994en}. This can be illustrated after introducing $n$-particle momentum correlations as
\begin{align}
C_n(p_1,\dots,p_n)\equiv\frac{N_n(p_1,\dots,p_n)}{N_1(p_1)\cdot\dots\cdot N_1(p_n)} \textnormal{ and}\label{e:cndef}\\
\lambda_n\equiv C_n(p_1=p_2=\dots=p_n)-1,
\end{align}
i.e. $\lambda_n$ is defined as the extrapolation of $C_n$ to zero relative momentum. It turns out that multi-particle Bose-Einstein correlation strengths are connected to the (partial) coherence of the fireball~\cite{Csorgo:1999sj}:
\begin{align}
  \lambda_2 =& \,f_c^2[(1-p_c)^2+2p_c(1-p_c)]\label{e:fcpc2}\\
\lambda_3 =& \,2f_c^3[(1-p_c)^3+3p_c(1-p_c)^2]+3f_c^2[(1-p_c)^2+2p_c(1-p_c)],\label{e:fcpc3}\\
\lambda_4 =& \,9f_c^4[(1-p_c)^4+4p_c(1-p_c)^3]+8f_c^3[(1-p_c)^3+3p_c(1-p_c)^2]
           +6f_c^2[(1-p_c)^2+2p_c(1-p_c)],\label{e:fcpc4}\\
&\textnormal{where}\quad p_c =  N_{\rm coherent}/N_{\rm core}.
\end{align}
This means that a simultaneous measurement of at least $\lambda_2$ and $\lambda_3$ in two- and
three-pion correlation functions offers the possibility of investigation of coherent pion production and determination $f_c$ and $p_c$~\cite{Csanad:2005nr,Csanad:2018vgk}. The above mentioned effects underline the importance of understanding the effects that may modify the pair momentum dependence of $\lambda_n$.
 
Note furthermore that based on Eqs.~\eqref{e:fcpc2}-\eqref{e:fcpc3}, one can define~\cite{Csanad:2018vgk} an $f_c$ independent combination $\kappa_3$:
\begin{align}
\kappa_3 = \frac{\lambda_3-3\lambda_2}{2\lambda_2^{3/2}} = \frac{1+p_c-2p_c^2}{(1+p_c)\sqrt{1-p_c^2}}.\label{e:kappa3def}
\end{align}
A departure of this parameter from unity, i.e. $\kappa_3\neq 1$, would mean that phenomena beyond the core-halo model have to be considered. One such phenomenon may be the above noted coherent pion production.

\section{Strength of multi-particle Bose-Einstein correlations}
\label{s:lambda23}

\begin{figure}
\begin{center}
\includegraphics[width=0.6\textwidth]{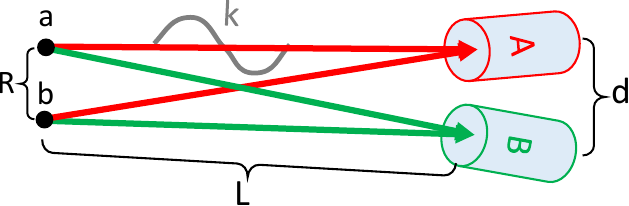}
\end{center}
\caption{A possible measurement setup with two point-like sources $a$ and $b$, and two detectors $A$ and $B$.}\label{f:setup}
\end{figure}

In this section we investigate the effect of the surrounding hadron gas on multi-particle Bose-Einstein correlations. In order to do so, let us recapitulate how the HBT effect can be explained in case of two point-like sources~\cite{Baym:1997ce}. As shown in Fig.~\ref{f:setup}, let there be two point like sources, $a$ and $b$, at a distance of $R$, emitting particles with wave functions $\Phi_a(r)$ and $\Phi_b(r)$ (in this scenario, clearly $f_c=1$ and $p_c=0$). Furthermore, let there be two detectors, $A$ and $B$, separated by $d$, and at an $L$ distance from the sources, with $d,R\ll L$. These detectors measure the total single-particle densities at their respective locations, $\Psi(r_A)$  and $\Psi(r_B)$. The coincidence amplitude is then $\Psi(r_A,r_B)$, and the correlation function is:
\begin{align}
C_{AB}=\frac{\langle |\Psi(r_A,r_B)|^2 \rangle}{\langle |\Psi(r_A)|^2 \rangle\langle |\Psi(r_B)|^2, \rangle}\label{e:CABdef}
\end{align}
where $\langle \cdot \rangle$ denotes a time average, which corresponds to a thermal average in case of thermal emission. Let us note furthermore that as shown in Fig.~\ref{f:setup}, the possible paths of the pair form a closed loop, and the fields enclosed within may modify the paths as well, such as in the Aharonov--Bohm effect. Here we restrict ourselves on a more classical effect, which is however similar in nature. In this section, the expression for the correlation function of Eq.~\ref{e:CABdef} is derived for thermally emitted particles in a random external field.

\subsection{Two-boson correlations with thermal emission}

Let the previously mentioned sources emit matter waves with wave number $k$ (which we identify here with momentum, in $\hbar=1$ \& $c=1$ units). For plane waves\footnote{The whole calculation is similar for spherical waves, and identical in end result.}, this can be expressed as:
\begin{align}
\Phi_a(r)&=\alpha e^{ik(r-r_a)+i\phi_a},\\
\Phi_b(r)&=\beta e^{ik(r-r_b)+i\phi_b},
\end{align}
where $\alpha$ and $\beta$ are the strength of the individual sources, $\phi_{a,b}$ are the (thermal, chaotic) phases of the waves emitted from each point, while $k$ is the momentum or wave number (vector) of the waves. These particles are detected in detectors $A$ and $B$, where the single- and two-particle wave-functions are
\begin{align}
\Psi(r_{A,B})&=\Phi_a(r_{A,B})+\Phi_b(r_{A,B})\\
\Psi(r_A,r_B)&=\frac{1}{\sqrt{2}}\left(\Phi_a(r_A)\Phi_b(r_B)+\Phi_a(r_B)\Phi_b(r_A)\right).
\end{align}
In an approximation where $d,R\ll L$, assuming uniformly distributed random thermal phases, for which average of factors like $e^{i(\phi_b-\phi_a)}$ is zero, the correlation function is then:
\begin{align}
C_{AB}(q)=\frac{\langle |\Psi(r_A,r_B)|^2 \rangle}{\langle |\Psi(r_A)|^2 \rangle\langle |\Psi(r_A)|^2 \rangle} -1 = \cos\frac{kRd}{L} = \cos(Rq),
\end{align}
where we utilised that $d/L$ is the angle between the two detectors, hence $\frac{kd}{L}=q$ is the momentum difference of the pair. At zero relative momentum, the correlation strength is:
\begin{align}
\lambda_2 = C_{AB}(0)=\left.\frac{\langle |\Psi(r_A,r_B)|^2 \rangle}{\langle |\Psi(r_A)|^2 \rangle\langle |\Psi(r_B)|^2 \rangle} \right|_{q=0} -1 = 1.
\end{align}

\subsection{Effect of a random field on two-particle correlation strengths}
Let us now investigate the case where random phase-shifts have to be applied based on the path of the particles (as it would be in case of a random field), i.e. instead of simply $\phi_a$ and $\phi_b$, phases like $\phi_{aA}$ and $\phi_{aB}$ (and similarly for $a \leftrightarrow b$) also appear. If $\phi=\phi_{aA}+\phi_{bB}-\phi_{aB}-\phi_{bA}$ is the total phase picked up through the closed loop represented by the possible paths of the pair (as shown in Fig.~\ref{f:setup}), then one gets
\begin{align}
C_{AB}(q)=\frac{\langle |\Psi(r_A,r_B)|^2 \rangle}{\langle |\Psi(r_A)|^2 \rangle\langle |\Psi(r_B)|^2 \rangle} -1 = 1+\cos(Rq+\phi).
\end{align}
If all $\phi_{xX}$ type of phases are independent, Gaussian random variables with width $\sigma$, then the four-term sum $\phi$ is also normally distributed, with a width of $2\sigma$. When averaging on these ``path-phases'' (i.e. taking the average $\langle \exp(i\phi)\rangle_{\phi}$ in addition to the averaging over the thermal phases) one obtains:
\begin{align}
C_{AB}(q) = \cos(R\Delta k)e^{-2\sigma^2},\quad\textnormal{and}\quad
\lambda_2 = C_{AB}(0) = e^{-2\sigma^2}.\label{e:l2sigma}
\end{align}

\subsection{Effect of a random field on multi-particle correlation strengths}
Three-particle correlations are defined by generalising the setup of Fig.~\ref{f:setup} to three sources ($a,b,c$) and three detectors ($A,B,C$). The correlation function is then:
\begin{align}
C_{ABC}(q) = \frac{\langle |\Psi(r_A,r_B,r_C)|^2 \rangle}{\langle |\Phi(r_A)|^2 \rangle\langle |\Phi(r_B)|^2 \rangle\rangle\langle |\Phi(r_C)|^2 \rangle},
\end{align}
where $q$ symbolises the momentum-differences within the triplet. The three-particle, symmetrised wave function is
\begin{align}
\Psi(r_A,r_B,r_C)=\frac{1}{\sqrt{6}}(
&\Phi_a(r_A)\Phi_b(r_B)\Phi_c(r_C)+
\Phi_a(r_B)\Phi_b(r_C)\Phi_c(r_A)+\\[-6pt]
&\Phi_a(r_C)\Phi_b(r_A)\Phi_c(r_B)+
\Phi_a(r_C)\Phi_b(r_B)\Phi_c(r_A)+\nonumber\\
&\Phi_a(r_A)\Phi_b(r_C)\Phi_c(r_B)+
\Phi_a(r_B)\Phi_b(r_A)\Phi_c(r_C)).\nonumber
\end{align}
The phase-averaged three-particle density has $6\times 6$ terms and would be very lengthy to write out, let us just mention here that it has the shape
\begin{align}
\langle |\Psi(r_A,r_B,r_C)|^2 \rangle&=\frac{1}{6}
\left[6+(\textnormal{30 cross-terms})\right].
\end{align}
Although the cross-terms are not explicitly written out here, it is important to note that all terms become unity at zero relative momenta when there are no path-related phases just the thermal emission-related phases. Therefore
\begin{align}
\lambda_3 = C_{ABC}(q) -1 =5.
\end{align}

If there are random phases picked up in different paths, then these enter in the three-particle wave function as terms like $i(\phi_{aA}+\phi_{bB}+\phi_{cC}))$ (and other permutations) in the exponents. In the end, out of the $6\times 6$ terms that appear in $C_{ABC}$, 6 are unity, 18 contain only pair-correlations (i.e. will contain four $\phi_{xX}$ terms and hence contribute with $\exp(-2\sigma^2)$), and there are 12 genuine 3-particle correlation-like terms (i.e. will contain six $\phi_{xX}$ phases, contributing with $\exp(-3\sigma^2)$). Summing all the random variables $\phi_{xX}$ and taking their averages yields:
\begin{align}
\lambda_3 = C_{ABC}(0)-1 = 3e^{-2\sigma^2}+2e^{-3\sigma^2}.\label{e:l3sigma}
\end{align}
Similarly, four-particle correlations $C_{ABCD}$ can also be calculated based on the 24-term symmetrised plane-wave. Then the $24\times 24$ terms in its absolute square can be counted (separately the ones with genuine four-particle correlations, the ones with three-particle correlations and the ones with pair-correlations), and and the result for the correlation strength at zero relative momentum is:
\begin{align}
\lambda_4 = C_{ABCD}(0)-1= 6e^{-2\sigma^2}+8e^{-3\sigma^2}+9e^{-4\sigma^2}.\label{e:l4sigma}
\end{align}
One can observe in Eqs.~\eqref{e:l2sigma}, \eqref{e:l3sigma} and \eqref{e:l4sigma} that $\exp(-\sigma^2)$ plays the role of the core fraction $f_c$ in these calculations. The reason for this is that in both cases---the core-halo model and this random field phenomenon---genuine pairs, triplets, etc. in the $n$-tuple give a fixed contribution to the correlation function, and counting these gives an identical result with $\exp(-\sigma^2)$ corresponding to $f_c$.

If $\sigma$ depends on momentum, then the above results yield a momentum-dependent suppression of two- and three-particle correlation strengths $\lambda_2$ and $\lambda_3$. As $\sigma$ represents the variance of path modification, it is reasonable to expect it to decrease with momentum. Instead of assuming a momentum-dependence however, in the next section we will estimate it via a simple toy model of a Hubble-expanding hadron gas and a probe particle.

\section{Quantitative estimations via a toy model}
\label{s:toymodel}
In case of heavy-ion collisions, after the hadronisation, hundreds of charged particles are created. These produce an electromagnetic field around the path of the given particle-pair investigated for measuring correlation functions. While---as also mentioned in the introduction---this may be interpreted as an Aharonov-Bohm effect when the path possibilities of the pair are drawn as a closed loop, a simpler picture is when just the path of each particle of the pair is modified by an additional phase due to the interaction with the hadron gas. The phase shift of the particles can be linked to the modification of the time of flight ($t_{\rm TOF}$) of the particle to the detector.

To quantify the phase shift, a simple model is set up, in which the time of flight shift can be calculated numerically. In this model, there are $N_{\rm c}$ normally distributed charged particles with a Gaussian width $R$, zero net-charge, and a 3D Hubble-type of flow profile.\footnote{Assuming Hubble-flow is a reasonable approximation for the evolution of the hadron gas created after freeze-out.} A probe particle of mass $m$ and initial momentum $p$ the lab-frame velocity is (in $c=1$ units) $v=p/\sqrt{p^2+m^2}$). Hence time that would be needed to reach a distance $d$ can be expressed as
\begin{align}
t_{\rm TOF}^{(0)}(d) = d\sqrt{1+\frac{m^2}{p^2}}.
\end{align}
This arrival time is modified by the electromagnetic interaction with the Hubble-expanding (and ever diluting) hadron gas, and a time-shift of
\begin{align}
\Delta t(d) = t_{\rm TOF}(d)-t_{\rm TOF}^{(0)}(d)
\end{align}
can be defined, where $t_{\rm TOF}^{(0)}$ shall be calculated based on the final momentum, modified by the electromagnetic scattering. While in an experimentally realised scenario, this difference is taken at a $d$ of several meters, in fact $\Delta t(d)$ converges much earlier. We found that $\Delta t(d)$ at a few tens of thousands of fm is sufficiently saturated. In the subsequent part of this manuscript, $\Delta t$ denotes this saturated (converged) value of $\Delta t(d)$. The time-shift is then connected to the phase-shift. In order to obtain that connection, let us express the phase shift as
\begin{align}
\phi=\Delta x k = \Delta t v \frac{p}{\hbar} = \Delta t \frac{p^2}{\hbar\sqrt{m^2+p^2}},
\end{align}
where we now explicitly write out $\hbar$ for the sake of numerical calculations. Based on this, the width of the time-shift distribution, $\sigma_t$, is related to the phase-shift distribution, $\sigma$, as:
\begin{align}
\sigma = \frac{\sigma_t p^2}{\hbar\sqrt{m^2+p^2}}.\label{e:sigma_sigmat}
\end{align}

\begin{figure}
\centering
\includegraphics[width=1\textwidth]{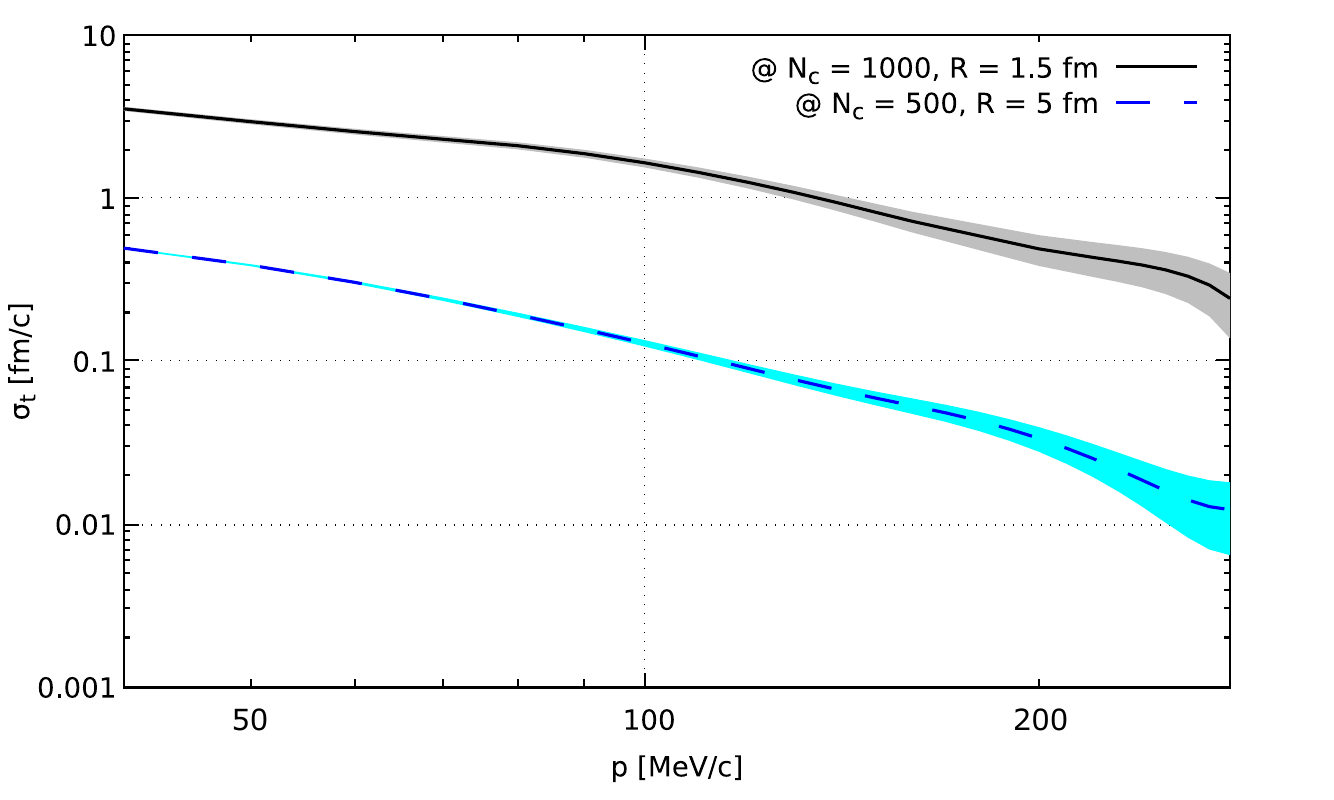}
\caption{The width of the time-shift ($\Delta t$) distribution, $\sigma_t$, as a function of $p$, for $N_{\rm c}=1000$ \& $R=1.5$ fm (solid black line) and for $N_{\rm c}=500$ \& $R=5$ fm (dashed blue line). Values at individual momenta were smoothed via a cubic spline to remove numerical fluctuations. The shaded regions show the average numerical variance.}\label{f:timewidth}
\end{figure}

From a study of 8 million event-by-event fluctuating hadron gas clouds, Gaussian distributions of time-shift $\Delta t$ emerged, with the width $\sigma_t$ depending on initial probe-particle momentum $p$.
The momentum dependence of $\sigma_t$, for two different values of both $N_{\rm c}$ and $R$, are shown in Fig.~\ref{f:timewidth}. Clearly $\sigma_t$ is decreasing with increasing probe particle momentum, as higher momentum particles are affected less by the same charges. Furthermore, larger charge density results in larger time-shifts, i.e. a broader time-shift distribution.

It is to be noted that the scenario where $N_{\rm c}=1000$ and $R=1.5$ fm is realistic only for extremely energetic $p+p$ collisions. The scenario where $N_{\rm c}=500$ and $R=5$ fm is more reasonable. The  dependence of $\sigma_t$ on $p$ in these two scenarios follows a similar pattern, with the absolute values differing, almost, by one order-of-magnitude. The larger absolute values obtained for $\sigma_t$ (with a smaller $N_{\rm c}$ and a larger $R$) have tangible effects on the two- and three-particle correlation strengths, as illustrated in the following section. 

\section{Results and discussion}\label{s:results}

\begin{figure}
\centering
\includegraphics[width=0.99\textwidth]{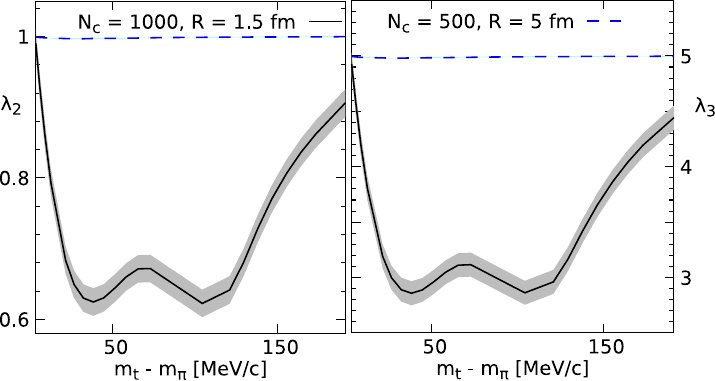}
\caption{\label{f:lambda} The $\lambda_2$ (right panel) and the $\lambda_3$ (left panel) values versus $m_t-m_\pi$ from model simulations (with $f_c=0$) at $N_{\rm c}=1000$ \& $R=1.5$ fm (solid, black line) and at $N_{\rm c}=500$ \& $R=5$ fm (dashed, blue line). Values at individual momenta were smoothed via a cubic spline to remove numerical fluctuations. The shaded regions show the average numerical variance (with negligible variance for the  $N_{\rm c}=500$ \& $R=5$ case, hence an extremely thin shaded band).}
\end{figure}

From the $\sigma_t(p)$ values shown in Fig.~\ref{f:timewidth}, $\sigma(p)$ may be calculated based on Eq.~\eqref{e:sigma_sigmat}. The resulting $\sigma$ values can then be substituted into Eqs.~\eqref{e:l2sigma}-\eqref{e:l3sigma}, and $\lambda_2$ and $\lambda_3$ can be plotted as functions of $m_t=\sqrt{K_t^2+m^2}$ (where $K=K_t$ at midrapidity, taken for comparability with experiment) as shown in Fig. \ref{f:lambda}. The modification of the $\lambda_{2,3}$ values as a function of $m_t$ is significant in case of the $N_c=1000$, $R=1.5$ fm scenario: there is a decrease from zero momentum (due to the momentum dependent factor in Eq.~\eqref{e:sigma_sigmat} being zero for $p=0$), and after an extremal range for $m_t-m_\pi$ values between approximately $40-130$ MeV$/c$, $\lambda_{2,3}$ both increase again to their default value, due to the diminishing $\sigma_t(p)$ values at large momenta. The modification is similar, but much smaller in magnitude (few \permil\:  changes) in case of the $N_c=500$, $R=5$ fm scenario. These results indicate that there may be cases where this effect has to be taken into account, especially at low pair transverse masses.

\section{Summary}
In summary, the idea was presented in this study suggests that random phases picked up throughout the flight towards the detector may distort the quantum-statistical correlations. As these correlations can be interpreted based on closed loops, formed by the possible paths of the involved particles, this effect is similar to the Aharonov--Bohm effect; where the additional phase is related to the fields enclosed by the loop. In this paper, the multi-particle correlation strengths were connected to the width of the phase-shift distribution. Subsequently, a simple model was presented where this width -- and hence the modification of correlation strengths -- could be estimated. This model is, by no means, complete or currently capable of delivering numerical results to be built upon. It shows nevertheless that, in principle, this phenomenon can have a non-negligible effect on quantum-statistical correlations. In conclusion, it is suggested to take the effect studied here into account when drawing conclusions on other physical phenomena related to correlation strengths.

\section{Acknowledgements}
This research was supported by the NKFIH grants FK-123842 and 2019-2.1.11-TÉT-2019-00080. M. Cs. is thankful for support of the Bolyai Scholarship of the Hungarian Academy of Sciences and the ÚNKP New National Excellence Program of the Hungarian Ministry of Innovation and Technology.

\end{document}